\documentclass[12pt]{iopart}
\usepackage{graphicx}

\usepackage{iopams}
\usepackage{graphicx}
\usepackage{dcolumn}
\usepackage{bm}
\usepackage[pdfstartview=FitH, bookmarksnumbered=true, bookmarksopen=true, colorlinks=true, pdfborder=001, citecolor=blue, linkcolor=blue, linktocpage=true] {hyperref}

\newcommand{\nn}{\nonumber}

\begin{document}

\title{Modulation-induced long-range magnon bound states in one-dimensional optical lattices}

\author{Wenjie Liu $^{1,2}$, Yongguan Ke $^{1,3}$, Bo Zhu $^{1}$, Chaohong Lee $^{1,2,4}$}

\address{$^{1}$Guangdong Provincial Key Laboratory of Quantum Metrology and Sensing $\&$ School of Physics and Astronomy, Sun Yat-Sen University (Zhuhai Campus), Zhuhai 519082, China}

\address{$^{2}$State Key Laboratory of Optoelectronic Materials and Technologies, Sun Yat-Sen University (Guangzhou Campus), Guangzhou 510275, China}

\address{$^{3}$Nonlinear Physics Centre, Research School of Physics, Canberra ACT 2601, Australia}

\address{$^{4}$Synergetic Innovation Center for Quantum Effects and Applications, Hunan Normal University, Changsha 410081, China}

\ead{lichaoh2@mail.sysu.edu.cn}

\begin{abstract}
Ultracold two-level atoms in optical lattices offer an excellent experimental platform to explore magnon excitations [Fukuhara \emph{et al} 2013 \emph{Nat. Phys.} \textbf{9}, 235; Fukuhara \emph{et al} 2013 \emph{Nature} \textbf{502}, 76].
Here, we investigate how gradient magnetic field and periodically modulated tunneling strength affect the two-magnon excitations in these ultracold atomic systems.
In the resonant condition where the driving frequency matches and smooths the potential bias, the system gains translational invariance in both space and time in the rotating frame, and thus we can develop a Floquet-Bloch band theory for two magnons.
We find a new kind of bound states with relative distance no less than two sites, apart from the conventional bound states with relative distance at one site,
which indicates the modulation-induced long-range interaction.
We analytically derive an effective Hamiltonian via the many-body perturbation theory for a deeper understanding of such novel bound states and explore the interplay between these two types of bound states.
Moreover, we propose to probe modulation-induced bound states via quantum walks.
Our study not only provides a scheme to form long-range magnon bound states, but also lays a cornerstone for engineering exotic quantum states in multi-particle Floquet systems.
\end{abstract}

\vspace{2pc}
\noindent{\it Keywords}: ultracold atoms in optical lattices, magnon bound states, Floquet-Bloch band

\date{\today}
\maketitle

\section{Introduction\label{Sec1}}

Periodic modulations in quantum systems have attracted tremendous interests and attentions in recent   years~\cite{MBukov2015,YKe2017,ZZLi2017,FrederikGorg2018,BWang2018,LLi2018,CHLeePRB2018}.
Periodic modulation not only provides a versatile tool to manipulate the quantum particles,
but also brings novel states of matter into the quantum systems~\cite{MSRudner2019}.
It has already been applied to control the hopping~\cite{VDalLago2015,BZhu2018}, band structure~\cite{WZheng2014,MHolthaus2015}, and quantized transport of a single particle~\cite{MLohse2016,SNakajima2016,YKe2016,OZilberberg2018,MLohse2018,SHu2019}.
Remarkably, artificial gauge field~\cite{NGoldman2014} and Floquet topological insulator~\cite{YTenenbaumKatan2013,JCayssol2013,MCRechtsman2013} have been realized by well-designed modulation protocols.
In principle, Floquet-Bloch theory of a single particle has been well developed to analyze the properties of periodically modulated systems~\cite{AGomezLeon2013}.
However, many-body effects induced by time modulations are more challenging and appealing.
Some novel Floquet many-body states, such as collective emission of matter-wave jets~\cite{LWClark2017}, have been realized in ultracold atomic systems.

Optical lattice of ultracold atoms, known as an excellent simulator, enables flexible engineering of dynamical modulations~\cite{AEckardt2017} as well as precise controlling of atom-atom interactions~\cite{AWidera2004,CGross2010}.
Owed to the excellent experimental techniques, it becomes possible to periodically modulate many-body systems with various methods.
Considering Hubbard-type models, periodic modulation of tunneling is used for engineering the interaction~\cite{EAnisimovas2015,LCardarelli2016}.
Modulation of both on-site energy and interaction is a route to create the nearest-neighbor (NN) interaction and density-assisted tunneling~\cite{HZZhao2019}.
Tunable three-body interaction was predicted to support fractional quantum hall states~\cite{CHLeePRL2018}.
Remarkably, density-dependent correlated tunneling~\cite{FMeinert2016} and density-dependent synthetic gauge fields~\cite{LWClark2018} have already been observed in optical lattices of cold atoms.

Optical lattice of ultracold atoms is also an ideal platform to simulate magnon excitations in a spin chain.
The dynamics of single magnon and two-magnon bound states in an undriven spin chain have been observed in ultracold atomic experiments~\cite{TFukuhara20131,TFukuhara20132}.
Motivated by these pioneer works, the studies of spin models have attracted broad interests in recent years~\cite{XQin2018,JAmatoGrill2019,IDimitrova2020}.
Interestingly, periodically driven gradient magnetic field in a spin chain is supposed to tune long-range interaction to short-range interaction~\cite{TELee2016}.
It is meaningful to engineer novel interactions in a time-modulated tilted spin chain with ultracold atoms.
Furthermore, it is also well worth exploring the correlation properties arising from the engineered interactions  and the way to probe the engineered interactions.

In this paper, we study two strongly interacting magnons in a spin chain under a gradient magnetic field and periodic modulation of hopping, which can be realized in a one-dimensional optical lattice of cold atoms.
When the potential bias between NN sites is equal to the driving frequency, a single magnon can resonantly tunnel to the neighboring sites.
We develop a Floquet-Bloch band theory for two magnons, because the system in the rotating frame is invariant when the two magnons are shifted as a whole by multiples of lattice constant and time period.
Apart from conventional bound states, we find that two magnons are bounded within two sites by calculating the magnon-magnon correlation, indicating effective next-nearest-neighbor (NNN) interactions.
An effective two-magnon model is obtained by using the many-body perturbation theory, which interprets the physical mechanism of modulation-induced long-range bound states.
The effective NNN interaction can be tuned to be positive or negative.
When the NN interaction equals to the modulation frequency, there exists a resonance between the bound states with relative one-site and two-site distances.
Considering the experimental realization, we also propose to probe the effective long-range two-magnon bound states via the quantum walks.

This paper is organized as follows.
In Sec.~\ref{Sec2}, we briefly describe the realization of driven two-magnon model.
In Sec.~\ref{Sec3}, we calculate the quasienergy spectrum as a function of NN interaction, by using the Floquet spectrum analysis in both time and frequency domains.
In Sec.~\ref{Sec4}, we analyze the Floquet-Bloch band, derive an effective two-magnon model for the modulation-induced long-range bound states, and show the resonance between two kinds of bound states.
In Sec.~\ref{Sec5}, we verify the long-range two-magnon bound states via the quantum walks.
In Sec.~\ref{Sec6}, we give a brief summary and discussion.

\section{Model \label{Sec2}}
Owed to current experiment techniques in manipulating ultracold atoms, single magnon and magnon bound states can be created by loading two-level atoms in a one-dimensional optical lattice~\cite{TFukuhara20131,TFukuhara20132}.
One may label the two hyperfine levels of atoms as spin up $|\uparrow\rangle$ and spin down $\left|\downarrow\right\rangle$.
Through applying a gradient magnetic field, spin up and spin down feel opposite potentials at the $l$-th site, $V_{l\uparrow}=-V_{l\downarrow}=lB$ with the gradient $B$. The Hamiltonian reads
\begin{eqnarray}\nn
\hat{H}_{\rm BH}=&-\sum_{<l,g>,\sigma=\uparrow,\downarrow} t_{\sigma}\hat{b}^{\dag}_{l\sigma}\hat{b}_{g\sigma}+\sum_{l\sigma}
\frac{U_{\sigma\sigma}}{2}\hat{n}_{l\sigma}(\hat{n}_{l\sigma}-1) \\
&+U_{\uparrow\downarrow}\sum_l\hat{n}_{l\uparrow}\hat{n}_{l\downarrow}+\sum_{l\sigma}V_{l\sigma}\hat{n}_{l\sigma}.
\end{eqnarray}
Here, $\sigma=\uparrow,\downarrow$, $\hat{b}^{\dag}_{l\sigma}$ ($\hat{b}_{l\sigma}$) is the particle creation (annihilation) operator,  $\hat{n}_{l\sigma}$  is the number operator, $t_\sigma$ is the hopping strength of $|\sigma\rangle$, and $U_{\sigma\sigma'}$ is the interaction strength between $|\sigma\rangle$ and $|\sigma'\rangle$.

In the Mott-insulator regime with one boson atom per site,
one can define the spin-1/2 operators $\hat{S}^\mathcal{I}_l(\mathcal{I}=x,y,z)$ as
\begin{eqnarray}\nn
\hat{S}^+_l&=\hat{b}^{\dag}_{l\uparrow}\hat{b}_{l\downarrow}, \\ \nn
\hat{S}^-_l&=\hat{b}^{\dag}_{l\downarrow}\hat{b}_{l\uparrow}, \\
\hat{S}^z_l&=(\hat{n}_{l\uparrow}-\hat{n}_{l\downarrow})/2.
\end{eqnarray}
Introducing the spin raising and lowering operators $\hat{S}^{\pm}_l=\hat{S}^x_l\pm i\hat{S}^y_l$ and the parameters
\begin{eqnarray}
J=\frac{2t_{\uparrow}t_{\downarrow}}{U_{\uparrow\downarrow}},\Delta=\frac{4t_{\uparrow}^2}{U_{\uparrow\uparrow}}+
\frac{4t^2_{\downarrow}}{U_{\downarrow\downarrow}}-\frac{2(t^2_{\uparrow}+t^2_{\downarrow})}{U_{\uparrow\downarrow}},
\end{eqnarray}
one can obtain a Heisenberg $XXZ$ chain,
\begin{eqnarray}\label{originalspinchain}
  \hat{H}=\sum_{l=-L}^L\big(J(t)\hat{S}^+_l\hat{S}^-_{l+1}+\rm{H.c.}+\Delta\it{\hat{S}^z_l\hat{S}^z_{l+\rm{1}}+lB\hat{S}^z_l} \big).
\end{eqnarray}
Here, the total chain length is $L_t=2L+1$. Below we focus on the case of periodically modulated parameter, $J(t)=(J_0+J_1 \cos(\omega t))/2$ with the d.c amplitude $J_0$, the a.c amplitude $J_1$ and the modulation frequency $\omega$.
Unlike the three-color modulations in the Fermi-Hubbard model~\cite{LCardarelli2016}, our Heisenberg $XXZ$ chain~(\ref{originalspinchain}) only involves a single frequency.
For simplicity, we choose the units of $J_0=\hbar=1$.
The longitudinal spin-exchange coupling $\Delta$ can be tuned by Feshbach resonance~\cite{AWidera2004,CGross2010} and we assume $\Delta \geq 0$.
The magnetic field gradient $B$ breaks the translational symmetry of the system.
The ground state is the fully ferromagnetic state $\left|\downarrow\downarrow\downarrow...\downarrow\right\rangle$ for a positive and sufficiently large $B$.
By flipping spins over the ground state $\left|\downarrow\downarrow\downarrow...\downarrow\right\rangle$, we obtain the excited states.
Once one considers the ground state $\left|\downarrow\downarrow\downarrow...\downarrow\right\rangle$ as a vacuum state, magnon can be regarded as the basic excitation around the ferromagnetic ground state.
Considering the mapping relations $\left|\downarrow\right\rangle\leftrightarrow \left|0\right\rangle, \left|\uparrow\right\rangle\leftrightarrow \left|1\right\rangle, \hat{S}^+_l\leftrightarrow\hat{a}^{\dag}_l, \hat{S}^-_l\leftrightarrow\hat{a}_l$, and $  \hat{S}^z_l\leftrightarrow\hat{n}_l-\frac{1}{2}$, it amounts to load magnons into the tilted optical lattice with periodically driven hopping rate and NN interaction, i.e.,
$\hat{H}=\sum_l \left(\frac{J(t)}{2}\hat{a}^{\dag}_l\hat{a}_{l+1}+{\rm H.c.}+\Delta\hat{n}_l\hat{n}_{l+1}+B l\hat{n}_l\right).$
$\hat{a}_{l}^{\dag}$ ($\hat{a}_l$) creates (annihilates) a magnon at site $l$ and satisfies the commutation relations of the hard-core bosons.
$\hat{n}_l=\hat{a}_{l}^{\dag}\hat{a}_l$ is the number operator.

By a unitary treatment $\hat{H}'=\hat{U}\hat{H}\hat{U}^{{\dag}}-i\hat{U}\frac{\partial}{\partial t}\hat{U}^{{\dag}}$ with $\hat{U}=\exp({i\sum_l lBt\hat{n}_l})$~\cite{ABuchleitner2003,ARKolovsky2003}, the driven magnon Hamiltonian in the rotating frame is given as,
\begin{eqnarray}\label{drivenmagnonmodel}
\hat{H}'=\sum_{l=-L}^L\big(J(t)e^{-iBt}\hat{a}^{\dag}_l\hat{a}_{l+1}+{\rm H.c.}\big)+\Delta\sum_{l=-L}^L\hat{n}_l\hat{n}_{l+1}.
\end{eqnarray}
If we impose periodic boundary condition, Hamiltonian~(\ref{drivenmagnonmodel}) preserves translational invariance, which is slightly different from Hamiltonian~(\ref{originalspinchain}) at the boundary.
We only consider the bulk properties where the tiny difference between Hamiltonian~(\ref{originalspinchain}) and~(\ref{drivenmagnonmodel}) takes no effect. For convenience, we focus on Hamiltonian~(\ref{drivenmagnonmodel}) with periodic boundary condition in the following paper.

\begin{figure}[htp]
\begin{center}
\includegraphics[width=0.8\textwidth]{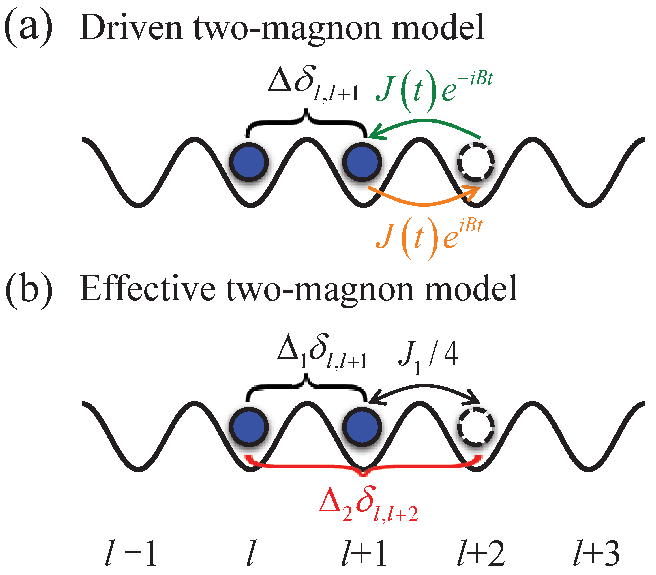}
\end{center}
\caption{(Color online)
(a) The driven two-magnon model~(\ref{drivenmagnonmodel}) with NN interaction and time-modulated hopping rate.
(b) The effective two-magnon model~(\ref{secondorderH}) with modulation-induced NNN interaction, modified NN interaction and reduced hopping rate.}
\label{fig:1}
\end{figure}

Since $[\hat{H'},\hat{n}]=0$ with $\hat{n}=\sum_l\hat{n}_l$, the total magnon number $\hat{n}$ is conserved.
This means that subspaces with different magnon numbers are decoupled.
To take magnon-magnon interaction into account,
two-magnon excitation is a fundamental object and gives enlightenment to multi-magnon excitations.
We aim to manipulate a NNN two-magnon bound state and explore its interplay with original NN bound state from both numerical and analytical perspectives.
Based upon a two-body ansatz related to center-of-mass and relative position, we concentrate our analysis on the subspace of two-magnon excitations.
The driven two-magnon model is schematically shown in figure~\ref{fig:1}(a).

In the absence of the gradient magnetic field and periodic driving
($B=J_1=0$), the spin-1/2 $XXZ$ chain occurs a quantum phase transition at $\Delta=J$.
For the two-magnon system, an isolated bound-state band and a continuum scattering band appear in the energy spectrum for $\Delta>J$,
while the bound-state band merges into the continuum one for $0<\Delta<J$.
After flipping two spins over the ground state with all spins downward, two magnons undergo Bloch oscillations in the presence of the gradient magnetic field~\cite{WLiu2019}.
Coherent delocalization occurs by introducing the resonant modulation $\omega=B$,
which is potentially applied for measuring the magnetic field gradient~\cite{MGTarallo2012,VVIvanov2008}.
In our paper, we only consider the resonant driven condition, i.e., $\omega=B$, where photon-assisted tunneling resonances may happen~\cite{NTeichmann2009,RMa2011}.

\section{Floquet spectrum analysis\label{Sec3}}

Only NN antiferromagnetic interaction ($\Delta>0$) is considered in the Hamiltonian~(\ref{drivenmagnonmodel}), while our conclusions can easily be extended to the corresponding ferromagnetic one ($\Delta<0$) due to the related symmetry analysis~\cite{WLiu2019}.
The interplay between NN interaction and time modulation makes it possible to induce long-range bound states.
To understand how the long-range bound states come from, we need to analyze the Floquet spectrum and the quasieigenstates.
There are two equivalent ways to calculate the Floquet-Bloch spectrum, one is in the time domain and the other is in the frequency domain.

In the following sections, we aim to study the modulation-induced NNN bound state and give an effective picture for its coexistence and competition with the original NN bound state.
In the subsection~\ref{Sec31} and \ref{Sec32}, we will respectively analyze how the Floquet spectrum changes with $\Delta$ in time and frequency domains in the high-frequency region.
However, in addition to the NNN bound states, our system can also be modulated to induce other bound states.
In~\ref{appendixA}, we give a brief discussion on the modulation-induced bound state with relative distance at three and four sites.

\begin{figure}[htp]
\begin{center}
\includegraphics[width=0.8\textwidth]{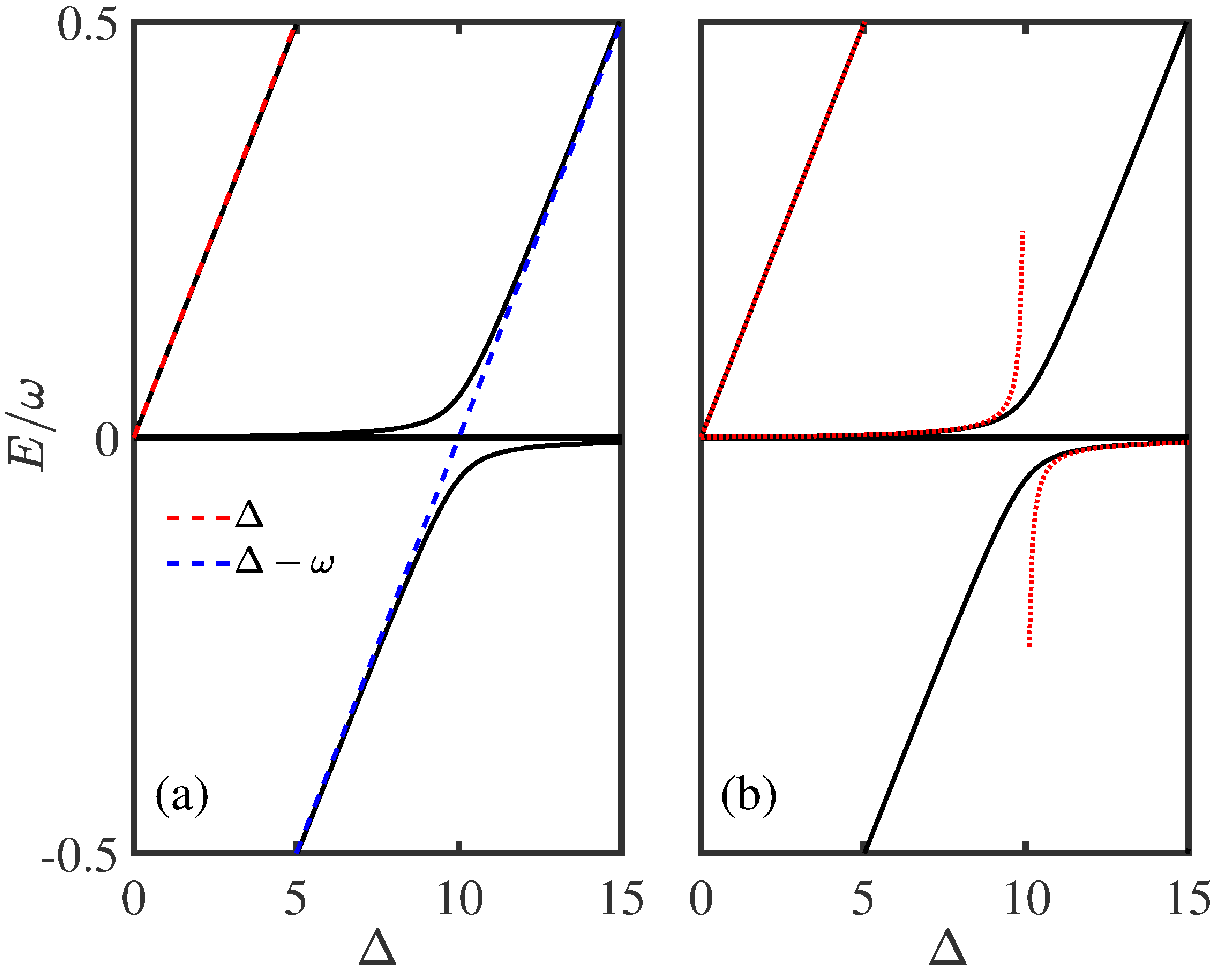}
\end{center}
\caption{(Color online)
(a) Quasienergy spectrum $E$ as a function of $\Delta$ given by the static effective Hamiltonian $\hat{H}_F$ (black solid lines). The red and blue dashed lines respectively represent $\Delta$ and $\Delta-\omega$ as a function of $\Delta$.
(b)  The quasienergy spectrums of the Floquet-Bloch lattice model~(\ref{FloquetBlochmodel}) (black solid lines) and the effective two-magnon model~(\ref{secondorderH}) (red dotted lines) for $F=5$.
The other parameters are chosen as $\omega=B=10$, $J_0=1$ and $J_1=0.01$.}
\label{fig:2}
\end{figure}

\subsection{Time-domain analysis \label{Sec31}}
The driven two-magnon Hamiltonian~(\ref{drivenmagnonmodel}) satisfies a discrete time translation symmetry, $\hat{H}(t+T)=\hat{H}(t)$ with a Floquet period $T=2\pi/\omega$.
We can define a time-evolution operator in one period as
\begin{equation}
\hat{U}_T=\hat{\mathcal{T}}\exp(-i\int^T_0\hat{H}'(t)dt)\equiv \exp(-i\hat{H}_F T),
\end{equation}
where the static effective Hamiltonian
\begin{eqnarray} \label{HamEffTime}
\hat{H}_F=\frac{i}{T}\log\hat{U}_T
\end{eqnarray}
governs the dynamics at stroboscopic time $nT$ ($n=1,\ 2,\ 3,...$).
Before analyzing the dynamics, it is helpful to solve the eigenvalue problem, $\hat{H}_F|u_n\rangle=E_n|u_n\rangle$,
where $E_n$ and $|u_n\rangle$ are quasienergy and  Floquet eigenstate, respectively.
$E_n$ can be restricted in the interval $[-\omega/2,\omega/2]$, which we term as the first Floquet-Brillouin zone.
The quasienergy has a period $\omega$ and consists of replicas of that in the first Floquet-Brillouin zone.
We calculate the quasienergy spectrum as a function of the NN interaction regarding the static effective Hamiltonian~(\ref{HamEffTime}) (see figure~\ref{fig:2}(a)).
The parameters are chosen as $\omega=B=10, J_0=1$ and $J_1=0.01$.
There is continuum flat band around zero energy, with eigenstates regarded as the scattering states.
When $J_0,\ J_1\ll \Delta$, two magnons at the NN sites approximate the bound states with energy $\sim \Delta$.
The red dashed line $\Delta$ as a function of $\Delta$ is added in figure~\ref{fig:2}(a), which is well fitting with an isolated band in quasienergy spectrum (black solid lines) of $\hat{H}_{F}$.
Thus, there is one isolated band $\sim\Delta$ corresponding to bound state with relative distance as $1$ in quasienergy spectrum of $\hat{H}_{F}$.
It becomes clearer that $\Delta-\omega$ (blue dashed line) is the replica of $\Delta$ (red dashed line).
Around the resonant condition $\Delta-\omega=0$, the bound-state band $\Delta-\omega$ (blue dashed line) is completely different from the quasienergy spectrum (black solid lines) of $\hat{H}_{F}$ and mixes with the continuum band, and two isolated bands appear in the quasienergy spectrum of $\hat{H}_{F}$ (see figure~\ref{fig:2}(a)).
Thus, the two isolated bands around $\Delta-\omega\approx 0$ must be a modulation-induced effect.

\subsection{Frequency-domain analysis \label{Sec32}}
We can equivalently analyze the quasienergy spectrum in the frequency domain.
The arbitrary two-magnon states can be expanded as $|\Psi\rangle=\sum_{l_1<l_2}\psi_{l_1l_2}|l_1l_2\rangle$, with probability amplitudes $\psi_{l_1l_2}=\langle\textbf{0}|\hat{a}_{l_2}\hat{a}_{l_1}|\Psi\rangle$ for one magnon at the $l_1$-th site and the other at the $l_2$-th site.
After substituting the two-magnon state in the Fock basis into the Schr\"{o}dinger equation
\begin{equation}\label{schrodingerequation}
i\frac{d}{dt}|\Psi(t)\rangle=\hat{H}'|\Psi(t)\rangle,
\end{equation}
the amplitude probability $\psi_{l_1l_2}(t)$ at instantaneous time $t$ satisfies
\begin{eqnarray} \label{ProbAmp} \nn
i\frac{\partial}{\partial t}\psi_{l_1l_2}(t)
&=&M(t)\big(\psi_{l_1,l_2+1}+\psi_{l_1+1,l_2}(t)\big)
+M^*(t)\big(\psi_{l_1,l_2-1}(t)+\psi_{l_1-1,l_2}(t)\big)\\
&&+\Delta\delta_{l_1,l_2\pm 1}\psi_{l_1l_2}(t)
\end{eqnarray}
with $M(t)={J_1}/{4}+{J_0}/{2}e^{-i\omega t}+{J_1}/{4}e^{-i2\omega t}$.
Considering the periodic boundary condition, we have $\psi_{l_1,l_2+L_t}=\psi_{l_1+L_t,l_2}=\psi_{l_1,l_2}$.
Due to the time periodicity, we can express the probability amplitudes $\psi_{l_1l_2}(t)$ by the Fourier components $\psi_{l_1l_2}(t)=e^{-iEt}\sum^{\infty}_{\chi=-\infty}e^{-i\chi \omega t}U_{l_1,l_2,\chi}$~\cite{BZhu2018, AGomezLeon2013}.
Then we have
\begin{eqnarray}\label{Floquetstate}
|\Psi(t)\rangle=e^{-iEt}\sum_{l_1<l_2,\chi}e^{-i\chi \omega t}U_{l_1,l_2,\chi}|l_1,l_2,\chi\rangle
\end{eqnarray}
where $\left|U_{l_1,l_2,\chi}\right|^2$ is the probability distributing in the Floquet state $|l_1,l_2,\chi\rangle$.
$\chi$ is the Floquet index taking values from $-\infty$ to $\infty$.
The Floquet state $|l_1,l_2,\chi\rangle$ intuitively represents two particles living in real space and Floquet space.
Due to the particle conservation, the two particles always share a common $\chi$, that is, the two particles transfer from $\chi$ to $\chi'$ as a bounded pair.
Replacing the state~(\ref{Floquetstate}) into the Eq.~(\ref{schrodingerequation}) and averaging it over one period,
we can obtain
\begin{eqnarray}\label{2DstaticHamiltonian}\nn
EU_{l_1,l_2,\chi}&=&(\Delta\delta_{l_1,l_2\pm 1}-\chi \omega)U_{l_1,l_2,\chi}
+\sum\limits_{q}\mathcal J_q\left(U_{l_1,l_2+1,{\chi-q}}+U_{l_1+1,l_2,{\chi-q}}\right. \\
&&+\left. U_{l_1,l_2-1,{\chi+q}}+U_{l_1-1,l_2,{\chi+q}}\right)
\end{eqnarray}
with $\mathcal J_{0}=\mathcal J_{2}={J_1}/{4}$ and $\mathcal J_1={J_0}/{2}$.
It is equal to a Floquet-Bloch lattice model,
\begin{equation}\label{FloquetBlochmodel}
\hat{H}_{FB}=\hat{H}_{FB}^0+\hat{H}_{FB}^1,
\end{equation}
with
\begin{eqnarray}
\hat{H}_{FB}^0&=&\sum\limits_{l_1,l_2,\chi}\big(\Delta\delta_{l_1,l_2\pm 1}-\chi \omega \big)|l_1,l_2,\chi\rangle\langle l_1,l_2,\chi|
\end{eqnarray}
and
\begin{eqnarray}\nn
\hat{H}_{FB}^1&=&\sum\limits_{l_1,l_2,\chi, q} \big(\mathcal J_q|l_1,l_2,\chi\rangle\langle l_1,l_2+1,\chi-q|+\rm{H.c.}\big) \\
&&+\sum\limits_{l_1,l_2,\chi, q} \big(\mathcal J_q|l_1,l_2,\chi\rangle\langle l_1+1,l_2,\chi-q|+\rm{H.c.}\big).
\end{eqnarray}
The Floquet index $\chi$ labels the extra dimension.
We establish an equivalence between a periodically driven two-magnon model~(\ref{drivenmagnonmodel}) and the Floquet-Bloch lattice model~(\ref{FloquetBlochmodel}).
The motion of two magnons on the periodically driven one-dimensional lattice in figure~\ref{fig:1}(a) is equivalent to the time-independent hopping dynamics of two magnon on a two-dimensional lattice model with a potential energy gradient along the $\chi$ direction.
Here, $\hat{H}_{FB}^0$ consists of the NN interaction and the potential gradient $-\chi\omega$ along $\chi$ direction.
$\hat{H}_{FB}^1$ consists of individual hopping of magnons in the same $\chi$ and pair-hopping between $\chi$ and $\chi'$ with $\chi'=\chi\pm 1,\ \chi \pm 2$.

Numerically, we can calculate the quasienergy spectrum by truncating the Floquet spaces ranging from $\chi=-f$ to $f$, and the total truncation number $F=2f+1$.
Generally, $f$ can be ranging from $-\infty$ to $\infty$.
In the high-frequency case, $\omega\gg J_0,\ J_1$, the energy barrier between $\chi$ and $\chi\pm1$ is so large that the two magnons tends to localize at a few $\chi$.
A small truncation number gives sufficiently exact results and enables us to apply the perturbation theory to obtain an effective Hamiltonian.
However, as the modulation frequency decreases, the wave function become spread over a larger range of $\chi$, and a larger truncation number is needed.
By choosing the same parameters as figure~\ref{fig:2}(a) and $F=5$, we also calculate the quasienergy spectrum with the change of $\Delta$ given by the Floquet-Bloch lattice model~(\ref{FloquetBlochmodel}) (black solid lines) in figure~\ref{fig:2}(b).
Compared with the quasienergy spectrum of $\hat{H}_F$ (black solid lines) in figure~\ref{fig:2}(a) and $\hat{H}_{FB}$ (black solid lines) in figure~\ref{fig:2}(b), they are almost the same. It means that these two methods are equivalent.
However, the analysis in the frequency domain is a more powerful method to understand the emergence of modulation-induced isolated bands more deeply.
Based on the analysis in the frequency domain, we will analyze the Bloch states under a periodic boundary condition far away and around the resonant parameter.

\section{Two-body Flqouet-Bloch band \label{Sec4}}
In this section, we derive a Floquet-Bloch lattice model which has cotranslational symmetry along the real space but violates such symmetry along the Floquet space due to the effective potential gradient $-\chi\omega$.
Thus, we can apply the many-body Bloch theorem in the real space and leave alone the Floquet index $\chi$.
Remarkably, we obtain two-body Floquet-Bloch band, a significant analysis of the many-body Bloch theorem to the periodically modulated system.
Away from the resonant condition $\Delta-\omega=0$, we analyze the isolated Floquet-Bloch bands and reveal the modulation-induced long-range two-magnon bound states.
Such process can also be perfectly captured by an effective Hamiltonian via the many-body perturbation theory.
Near the resonant condition, we find the hybrid of two kinds of bound states.

\begin{figure}[htp]
\begin{center}
\includegraphics[width=0.7\textwidth]{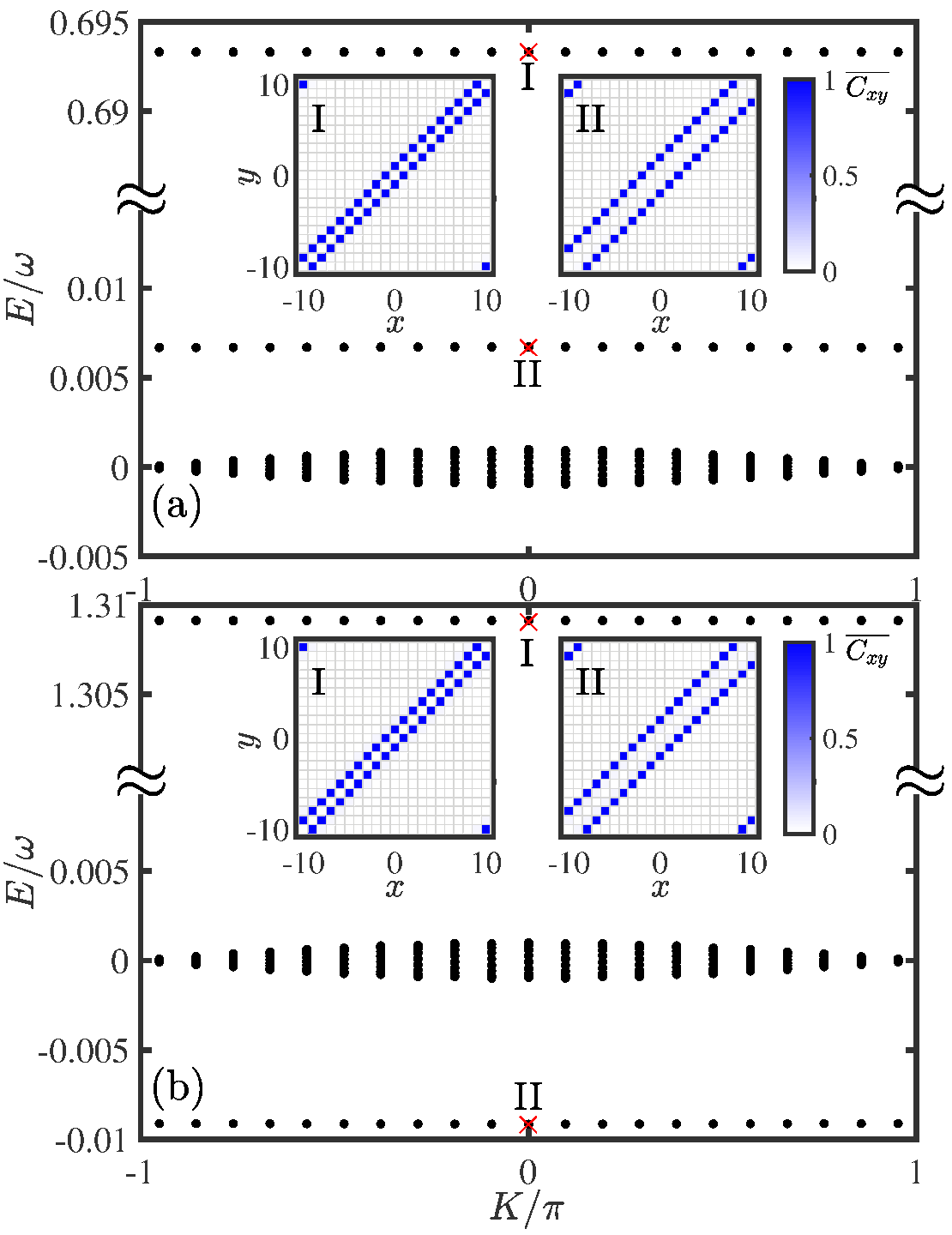}
\end{center}
\caption{(Color online)
Quasienergy spectrum $E$ vs. center-of-mass momentum $K$ away from the resonant condition for (a) $\Delta-\omega=-3$ and (b) $\Delta-\omega=3$.
The insets $\rm{\uppercase\expandafter{\romannumeral1}}$ and $\rm{\uppercase\expandafter{\romannumeral2}}$ respectively describe the normalized magnon-magnon correlations $\overline{C_{xy}}=C_{xy}/C^{max}_{xy}$ of Flqouet states labeled in the isolated bands.
The other parameters are chosen as $\omega=B=10$, $J_0=1$ and $J_1=0.01$.}
\label{fig:3}
\end{figure}

\subsection{Modulation-induced long-range two-magnon bound states \label{Sec41}}
Imposing periodic boundary condition, the Floquet-Bloch lattice model~(\ref{FloquetBlochmodel}) is invariant by shifting the two magnons as a whole in the real space.
Naturally, we introduce the center-of-mass and relative positions $R=(l_1+l_2)/2$ and $r=l_1-l_2$, respectively.
The center-of-mass quasimomentum $K$ is a conserved quantity.
According to the many-body Bloch theorem, the wave-function is a Bloch wave along the coordinate of center-of-mass position, i.e., $U_{l_1,l_2,\chi}=e^{iKR}\phi_{\chi}(r)$ where $\phi_{\chi}(r)$ is the amplitude depending on the relative position $r$ and the Floquet position $\chi$.
It is difficult to assume such ansatz for three or more magnon excitations.

Substituting the above ansatz into Eq.~(\ref{FloquetBlochmodel}), the amplitude $\phi_{\chi}(r)$ satisfies the following eigenequation in the quasimomentum space,
\begin{eqnarray}\label{eigenequationinKspace}
E\phi_\chi(r)&=&\sum\limits_{q}\mathcal J^{K}_q\big(\phi_{\chi-q}(r-1)+\phi_{\chi-q}(r+1)\big)
+\big(\Delta\delta_{r,\pm 1}-\chi \omega\big)\phi_\chi(r)
\end{eqnarray}
with $\mathcal J^{K}_0={J_1}/{2}\cos({K}/{2})$, $\mathcal J^{K}_{\pm1}={J_0}/{2}e^{\pm{iK}/{2}}$ and $\mathcal J^{K}_{\pm2}={J_1}/{4}e^{\pm{iK}/{2}}$.
Under the periodic boundary condition, we find $e^{iKL_t}=1$ and
$\phi_{\chi}(r+L_t)=e^{iKL_t/2}\phi_{\chi}(r)$ with $K=2\pi\alpha/L_t$ for $\alpha=-L,-L+1,\ldots,L$.
Moreover, since the magnons are hard-core bosons, we have $\phi_{\chi}(0)=0$ and $\phi_{\chi}(r)=\phi_{\chi}(-r)$.

Solving the above eigenequation~(\ref{eigenequationinKspace}), we can obtain the two-body Floquet-Bloch bands vs. $K$ in figure~\ref{fig:3} (more results for longer-range bound states are shown in~\ref{appendixA}).
The parameters are chosen as $\omega=B=10$, $J_0=1$, $J_1=0.01$, and $\Delta$ are chosen at the left- and right-hand sides of the resonant point as $\Delta=7,\ 13$ for figures~\ref{fig:3}(a) and (b), respectively.
At the left-hand side of the resonant point, there are two isolated bands above a continuum band which ranges from $-|J_1|$ to $|J_1|$ (see figure~\ref{fig:3}(a)).
This is qualitatively different from a conventional two-band energy spectrum of two-magnon excitations in NN interaction quantum spin chains without the periodic driving~\cite{XQin2014}.
The extra band appears with quite a similar structure to the bound-state band arising from the NN interaction.
While at the right-hand side of the resonant point, the two isolated bands sandwich the continuum band (see figure~\ref{fig:3}(b)).
It will be clear later that the difference between figures~\ref{fig:3}(a) and (b) relates to the positive and negative values of the effective interaction.
A suitable choice of parameters results in a negative modulation-induced bound band below the continuum band.

To better understand the band structure, it is instructive to calculate the magnon-magnon correlations of corresponding Flqouet states labeled $\rm{\uppercase\expandafter{\romannumeral1}}$ and $\rm{\uppercase\expandafter{\romannumeral2}}$ in figure~\ref{fig:3}.
The magnon-magnon correlation is defined as $C_{xy}=\langle\Psi(T)|\hat{a}^{\dag}_{x}\hat{a}^{\dag}_{y} \hat{a}_{y}\hat{a}_{x}|\Psi(T)\rangle$ where $|\Psi(T)\rangle$ is the Floquet state with a given quasimomentum $K$.
$x$ and $y$ take values from $-L$ to $L$.
The magnon-magnon correlations at two specific lines $x=y\pm d$ in the $(x,y)$ plane serve as a sensitive evidence of the two-magnon bound states, where $d$ depends on the specific magnon-magnon interactions.
For example, due to the NN interaction, the Floquet state marked with $\rm{\uppercase\expandafter{\romannumeral1}}$ is well distributed among the minor-diagonal lines $x=y\pm 1$ of the magnon-magnon correlations in the inset $\rm{\uppercase\expandafter{\romannumeral1}}$ of figure~\ref{fig:3}.
While for the Floquet state marked with $\rm{\uppercase\expandafter{\romannumeral2}}$, the inset $\rm{\uppercase\expandafter{\romannumeral2}}$ of figure~\ref{fig:3} shows the magnon-magnon correlations are mainly distributed in the next-minor-diagonal lines.
The type $\rm{\uppercase\expandafter{\romannumeral2}}$ Floquet state indicates the existence of effective long-range bound states.
The correlation properties of the other Floquet states in these two bands are similar to their Floquet states with quasimomentum $K=0$.
They are characteristic signatures of two types of bound-state bands, respectively derived from the NN interaction and NNN interaction.
Thus, we can claim the existence of modulation-induced long-range bound states.

\subsection{Effective two-magnon model\label{Sec42}}

Below particular attention is paid to attain an effective two-magnon model for understanding and interpreting the origination of the modulation-induced long-range bound states.
Given the perturbation conditions $|\Delta-\omega|\gg J_0/2$ and $|\Delta-2\omega|\gg J_1/4$,
we divide the Floquet-Bloch lattice model~(\ref{FloquetBlochmodel}) into two parts, $\hat{H}_{FB}^0$ as a dominate term and $\hat{H}_{FB}^1$ as a perturbed term.
In the high-frequency region, $\omega\gg J_0,\ J_1$, it is sufficient to just take into account the five $\chi$ with $\chi=0,\pm1,\pm2$.
The unperturbed term $\hat{H}_{FB}^0$ is separated into two subspaces $\mathcal{U}$ and $\mathcal{V}$.
The subspace $\mathcal{U}$ includes two kinds of states:
(i) $E_P=\Delta$ for the states $\{|l,l+1,0\rangle\}$ with $ -L\leq l\leq L$ and
(ii) $E_P=0$ for the states $\{|l_1,l_2,0\rangle\}$ with $l_1\neq l_2-1, -L\leq l_1<l_2\leq L$.
The complementary subspace $\mathcal{V}$ consists of
(iii) $E_S=\Delta-\chi \omega$ for the states $\{|l,l+1,\chi\rangle\}$ with $-L\leq l\leq L$ and
(iv) $E_S=-\chi \omega$ for the states $\{|l_1,l_2,\chi\rangle\}$ with $l_1\neq l_2\pm1, -L\leq l_1<l_2\leq L$ for $\chi=\pm1,\pm2$.
The project operators are defined as $\hat{P}=\sum_{l_1l_2}|l_1,l_2,0\rangle\langle l_1,l_2,0|$ onto $\mathcal{U}$ and $\hat{S}=1-\hat{P}$ onto $\mathcal{V}$.
We calculate the effective two-magnon model $\hat{H}_{Eff}=\hat{h}_0+\hat{h}_1+\hat{h}_2$ via a perturbative expansion up to the second order.
In the lowest order and first order, we have
\begin{eqnarray}
\hat{h}_0=E_P\hat{P}=\Delta\sum\limits_{l}|l,l+1,0\rangle\langle l,l+1,0|
\end{eqnarray}
and
\begin{eqnarray}
\hat{h}_1&=\hat{P}\hat{H}_{FB}^1\hat{P} \cr\cr
=&\frac{J_1}{4}\sum\limits_{l_1l_2}\big(|l_1,l_2,0\rangle\big(\langle l_1,l_2+1,0|+\langle  l_1+1,l_2,0|\big)+\rm{H.c.}\big),
\end{eqnarray}
which respectively retain the original NN interaction $\Delta$ and NN tunneling in $\chi=0$.
For simplicity, we respectively label the eigenstates and eigenvalues of the unperturbed term $\hat{H}_{FB}^0$ as $\{|i\rangle\}$ and $E_i$,
and $\{|i_p\rangle\}$ represent states in the subspace $\mathcal{U}$.
The second-order effective Hamiltonian reads
\begin{eqnarray}\nn
\hat{h}_2&=&\sum\limits_{i_1i_3\in i_p,i_2\notin i_p}\frac{J^{Eff}_{i_1i_2i_3}}{2}
|i_1\rangle\langle i_1|\hat{H}_{FB}^1|i_2\rangle\langle i_2|\hat{H}_{FB}^1|i_3\rangle\langle i_3| \\
&=&-\Delta_2\sum\limits_{l}\big(|l,l+1,0\rangle\langle l,l+1,0|-|l,l+2,0\rangle\langle l,l+2,0|\big)
\end{eqnarray}
with $J^{Eff}_{i_1i_2i_3}=\big({1}/{(E_{i_1}-E_{i_2})}+{1}/{(E_{i_3}-E_{i_2})}\big)$ and $\Delta_2={J^2_0\Delta}/{[2(\omega^2-\Delta^2)]}+{J^2_1\Delta}/{[8(4\omega^2-\Delta^2)]}$.
Interestingly, the second-order process not only contributes to the NNN interaction, but modifies the original NN interaction.
Since the high-order term is much smaller than the second-order one in the perturbation parameter region, it allows us to ignore the results beyond the second-order term.

By means of the perturbation theory~\cite{SBravyi2011,MTakahashi1977,CLee2004}, the effective two-magnon model up to second order in $\chi=0$ is given as
\begin{eqnarray}\label{secondorderH}
\hat{H}_{Eff}&=&\frac{J_1}{4}\sum\limits_{l}\big(\hat{a}^{\dag}_l\hat{a}_{l+1}+{\rm H.c.}\big)+\sum\limits_{s=1,2;l}\Delta_s\hat{n}_l\hat{n}_{l+s}
\end{eqnarray}
with $\Delta_1=\Delta-\Delta_2$.
$\sum_l\hat{n}_l=2$ restricts $\hat{H}_{Eff}$ to the two-magnon sector.
The effective two-magnon model~(\ref{secondorderH}) can be schematically described in figure~\ref{fig:1}(b) with the renormalized parameters.
The effective NNN interaction comes from the asymmetric pathways between absorbing and consequently emitting phonons,  $\{|l_2-l_1=2,0\rangle \to |l_2-l_1=1,+\chi \rangle\to|l_2-l_1=2,0\rangle, \chi\neq0\}$ and the inverse process, $\{|l_2-l_1=2,0\rangle \to |l_2-l_1=1,-\chi\rangle\to|l_2-l_1=2,0\rangle, \chi\neq0\}$.
This is the origin of NNN interactions in a periodically modulated interacting system.
The hopping rate is reduced to $J_1/4$.
It must be pointed out that modulation amplitude $J_1$ determines the width of the continuum band $[-|J_1|, |J_1|]$.
Once $J_1$ is sufficiently large, the NNN-interaction bound-state band may not be enough to completely separate from the continuum band.
With the effective two-magnon model~(\ref{secondorderH}), we can systematically create and control the NNN interaction.
The long-range two-magnon bound states arising from the modulation-induced NNN interaction constitutes the central idea of our paper.

To show the concreteness, we compare the quasienergy spectrum given by the effective two-magnon model~(\ref{secondorderH}) and that given by $\hat{H}_{FB}$, respectively see the red dotted and black solid lines in figure~\ref{fig:2}(b).
It is clear that the isolated bands are well consistent in the perturbation parameter regime $|\Delta-\omega|\gg{J_0}/{2}$ and $|\Delta-2\omega|\gg{J_1}/{4}$.
The perturbation conditions mean the energy gap between subspaces $\hat{P}$ and $\hat{S}$ should be much larger than their tunneling rate.
This energy gap decreases as $\Delta$ approaches to $\omega$, so that the perturbation condition is no longer satisfied and the effective two-magnon model~(\ref{secondorderH}) is invalid.
As shown in figure~\ref{fig:2}(b) (red dotted lines), the isolated bound band of $\hat{H}_{Eff}$ is no longer fitting with the modulation-induced band of $\hat{H}_{FB}$ when the NN interaction $\Delta$ nears the modulation frequency $\omega$.
But their continuum bands are always well consistent.
In order to clearly clarify the modulation-induced bound band, we just plot the continuum band of $\hat{H}_{FB}$ in figure~\ref{fig:2}(b).

\subsection{Resonance between two types of bound states\label{Sec43}}
\begin{figure}[htp]
\begin{center}
\includegraphics[width=0.7\textwidth]{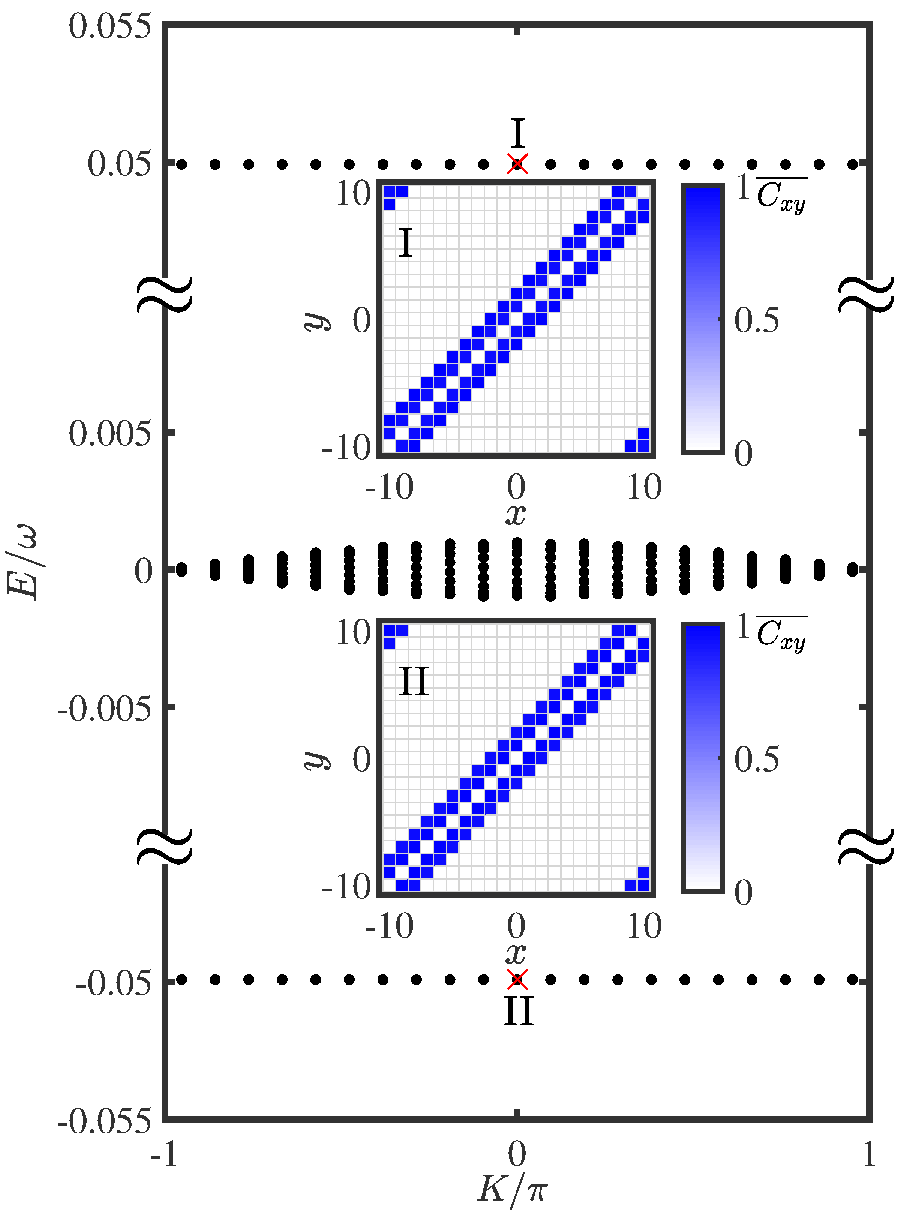}
\end{center}
\caption{(Color online)
Quasienergy spectrum $E$ vs. $K$ in the resonant condition $\Delta=\omega$.
The insets $\rm{\uppercase\expandafter{\romannumeral1}}$ and $\rm{\uppercase\expandafter{\romannumeral2}}$ respectively describe the normalized magnon-magnon correlations $\overline{C_{xy}}=C_{xy}/C^{max}_{xy}$ of Floquet states labeled in the top and bottom band in the quasienergy spectrum.
The other parameters are chosen as $\omega=B=10$, $J_0=1$ and $J_1=0.01$.}
\label{fig:4}
\end{figure}

However, the effective Hamiltonian~(\ref{secondorderH}) becomes invalid around the resonant point $\Delta=\omega$.
To understand what happens in the resonant condition,
we calculate the band structure and magnon-magnon correlations via the Floquet spectrum analysis in the frequency domain (see figure~\ref{fig:4}).
The truncated Floquet space is limited in $\chi=0,\pm1,\pm2$.

For $\hat{H}_{FB}^0$, its eigenstates  $\{|l,l+1,\chi\rangle: -L\leq l\leq L\}$ and  $\{|l_1,l_2,\chi-1\rangle: l_1\neq l_2-1, -L\leq l_1<l_2\leq L\}$ become degenerate with the energy difference $\Delta-\omega=0$.
Once $\hat{H}_{FB}^1$ is added, this energy degeneracy will be broken, and these states are not longer eigenstates of the full Hamiltonian~(\ref{FloquetBlochmodel}).
Two types of bound states $|l_2-l_1=1,\chi\rangle$ and $|l_2-l_1=2,\chi-1\rangle$ are coupled by a first-order process with the tunneling rate $J_0/2$, while state $|l_2-l_1=1,\chi\rangle$ couples state $\{|l_2-l_1>2,\chi-1\rangle: l_1\neq l_2-1,l_2-2\}$ by second-order or even higher-order term, which can be neglected in the high-frequency region.
Here we are able to just consider the first-order tunneling process between two types of bound states $|l,l+1,\chi\rangle$ and $|l,l+2,\chi-1\rangle$, then the corresponding states of the system~(\ref{FloquetBlochmodel}) can be written as a superposition $(|l,l+1,\chi\rangle\pm|l,l+2,\chi-1\rangle)/\sqrt{2}$ and the corresponding energies are $E=E_0\pm J_0/2$.

Now we focus on the zeroth Floquet-Brillouin zone $E\in[-\omega/2,\omega/2]$, there is no energy gap between bound states $|1\rangle=|l,l+1,1\rangle$ and $|2\rangle=|l,l+2,0\rangle$ with energy $E_0=0$.
In terms of a basis $\{|1\rangle, |2\rangle\}$, we build a two-level system with a degenerate energy $E_0$.
Once the first-order tunneling process $J_0/2$ is added, states $|1\rangle$ and $|2\rangle$ are not eigenstates of the full system whose eigenstates follow $|\pm\rangle=(|1\rangle\pm|2\rangle)/{\sqrt{2}}$ and eigenvalues are $E_0\pm J_0/2$.
The energy bands at the top and at the bottom are almost symmetrical with respect to $E_0=0$, and the energy gap is $J_0$ (see figure~\ref{fig:4}).
The parameters are chosen as $J_0=1$, $J_1=0.01$ and $\Delta=\omega=10$, in the high-frequency region $\omega\gg J_0,\ J_1$.
We respectively mark the two Floquet states as $\rm{\uppercase\expandafter{\romannumeral1}}$ and $\rm{\uppercase\expandafter{\romannumeral2}}$ in the quasienergy spectrum and analyze their correlated properties in the insets $\rm{\uppercase\expandafter{\romannumeral1}}$ and $\rm{\uppercase\expandafter{\romannumeral2}}$ of figure~\ref{fig:4}.
The top band and bottom one behave almost the same, that is, the magnon-magnon correlation is basically equal probability distribution on the minor-diagonal and next-minor-diagonal lines.
These numerical results completely follow the related theoretical analysis.

There exists a process in which, with the increasing of $\Delta$, a bound-state band caused by the NNN interaction in $\chi=0$ emerges from the continuum band.
When $\Delta$ nears $\omega$, it gradually mixes with the bound-state band caused by the NN interaction in $\chi=1$.
The NN-interaction bound-state band in $\chi=1$ plays a dominant role
until $\Delta$ is large enough.
Similarly, with the increasing of $\Delta$, an isolated NN-interaction bound-state band of $\chi=1$ appears below the continuum band, is mixed with the NNN-interaction one in $\chi=0$ around the resonant point $\Delta=\omega$, and finally turns to the one of the NNN interaction in $\chi=0$ for a sufficiently large $\Delta$.

\section{Probing long-range two-magnon bound states \label{Sec5}}
Two-magnon quantum walk provides an excellent method to probe the modulation-induced long-range two-magnon bound states.
It is worth to numerically simulate the dynamics of two strongly correlated magnons, initially localizing on the sites $l=-1$ and $l=1$, which can be prepared by flipping two NNN spins from a saturated ferromagnetic state with all spins downward $\left|\downarrow\downarrow...\downarrow\right\rangle$.
To investigate the dynamics of magnons initially locating  in the bulk of the spin chain, we resort to numerically solve the time-dependent Schr\"{o}dinger equation.
Starting from the driven spin-1/2 Heisenberg $XXZ$ chain under a gradient magnetic field~(\ref{originalspinchain}), the time evolution of an arbitrary two-magnon state $|\psi(t)\rangle$ obeys the time-dependent Schr\"{o}dinger equation $i\frac{d}{dt}|\psi(t)\rangle=\hat{H}|\psi(t)\rangle$,
where $|\psi(t)\rangle=\sum_{l_1<l_2}\psi_{l_1l_2}(t)|l_1l_2\rangle$ and the probability amplitudes $\psi_{l_1l_2}(t)=\langle\textbf{0}|\hat{S}^-_{l_2}\hat{S}^-_{l_1}|\psi(t)\rangle$.
\begin{figure}[htp]
\begin{center}
\includegraphics[width=0.8\textwidth]{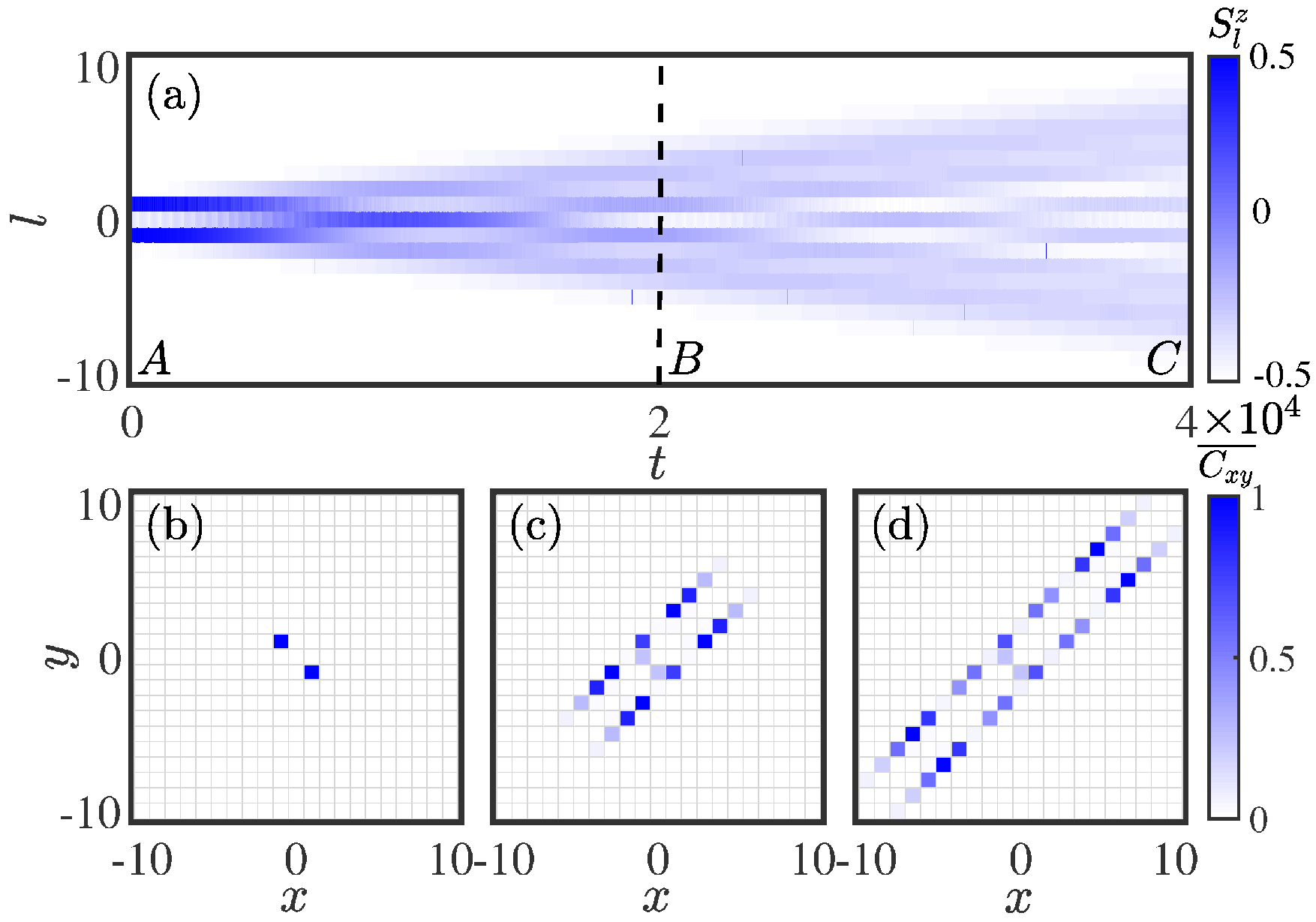}
\end{center}
\caption{(Color online) Quantum walks of two magnons in the driven Heisenberg $XXZ$ chain under a gradient magnetic field~(\ref{originalspinchain}).
(a) is the time evolution of spin distributions $S^z_{l}(t)$.
(b), (c) and (d) are the normalized magnon-magnon correlations $\overline{C_{xy}}=C_{xy}/C^{max}_{xy}$ for different moments marked as A, B and C in (a).
At the initial moment A ($t=0$), two magnons are prepared in $|-1,1\rangle$, so one can see two points in (-1,1) and (1,-1) in (b).
(c) and (d) reveal that, in the course of the time evolution, two magnons mainly move along the next-minor-diagonal lines of the magnon-magnon correlation.
The parameters are chosen as $\Delta=7$, $\omega=B=10$, $J_0=1$ and $J_1=0.01$.}
\label{fig:5}
\end{figure}
With the instantaneous wave function $|\psi(t)\rangle$, we trace out the time-dependent spin distributions $S^z_{l}(t)=\langle\psi(t)|\hat{S}^z_{l}|\psi(t)\rangle$
and the instantaneous magnon-magnon correlations $C_{xy}(t)=\langle\psi(t)|\hat{S}^+_{x}\hat{S}^+_{y} \hat{S}^-_{y}\hat{S}^-_{x}|\psi(t)\rangle$
for different moments, where $l$, $x$ and $y$ take values from $-L$ to $L$.

Figure~\ref{fig:5} displays the time evolution of spin distributions and the corresponding instantaneous magnon-magnon correlations at three different moments for $J_0=1$, $J_1=0.01$, $\Delta=7$, $\omega=B=10$, and $L_t=21$, with the periodic boundary condition.
Figure~\ref{fig:5}(a) shows a light cone in the time evolution of spin distributions before the two magnons collide with the boundaries.
We timely cut off the evolution of spin distributions to avoid the boundary effects and analyze its magnon-magnon correlations, marking A, B and C in figure~\ref{fig:5}(a).
As shown in figures~\ref{fig:5}(b), (c) and (d), almost all of the magnons are distributed among the next-minor-diagonal lines in the magnon-magnon correlation.
These results follow from the circumstance that two strongly NNN interacting magnons initially occupying the NNN sites form a bound pair and tunnel together on the spin chain.
These time evolved results are in agreement with our analytical predictions.
The two-magnon quantum walks pave the way for experimentally verifying whether the modulation-induced long-range bound states exist or not.

\section{Summary and Discussion\label{Sec6}}
Based on the state-of-art techniques of ultracold atomic experiments, our model can be experimentally simulated with ultracold two-level atoms in one-dimensional optical lattices.
We have presented a scheme for generating tunable long-range magnon bound states by introducing a gradient magnetic field and time-modulated hopping into the ultracold atomic systems.
Taking advantage of the flexible tunability,
desired long-range bound states can be engineered by changing the modulated parameters and the interaction strength $\Delta$ tuned by Feshbach resonance.
The prerequisite lies in the modulation frequency $\omega$ is equal to the magnetic field gradient $B$, which enables a single magnon to tunnel resonantly onto its neighbors.
It can be understood as the phonon-assisted tunneling among the spin chain.
We not only analyze the correlation properties of modulation-induced long-range bound states, but also explore the interplay between the original and modulation-induced bound states with the changing of NN interaction.
Further, we obtain an effective two-magnon model via a many-body perturbation theory to interpret the origin and condition for the remarkable long-range bound states.
The effective two-magnon model becomes invalid near the resonant point $\Delta=\omega$, where resonance between the two types of bound states happens.
In addition, two-magnon quantum walks can be used as an experimental verification of modulation-induced long-range bound states.
Our scheme is not limited to two-magnon systems, and also gives insights into engineering novel states of matter of multi-particle Floquet systems.

A few promising questions remain open and motivate further investigations.
For example,
this proposed scheme has potential to realize the fractional topological states which exist in the one-dimensional superlattice with the dipole-dipole interactions~\cite{ZXu2013}.
By introducing the modulation-induced NNN interaction, the competitive relation between two kinds of bound states may bring different topological effects in periodically modulated NN-interaction spin chain.
In the whole text, we focus on the analysis in the high-frequency region, while more abundant phenomenons may arise in the
low-frequency one.
%

\section{Acknowledgements}
This work is supported by the Key-Area Research and Development Program of GuangDong Province under Grants No. 2019B030330001, the National Natural Science Foundation of China (NNSFC) under Grants [No. 11874434, No. 11574405], and the Science and Technology Program of Guangzhou (China) under Grants No. 201904020024. Y.K. is partially supported by the Office of China Postdoctoral Council (Grant No. 20180052), the National Natural Science Foundation of China (Grant No. 11904419), and the Australian Research Council (DP200101168).
\appendix

\section{Longer-range bound states} \label{appendixA}

\begin{figure}[!htp]
\center
\includegraphics[width=0.8\textwidth]{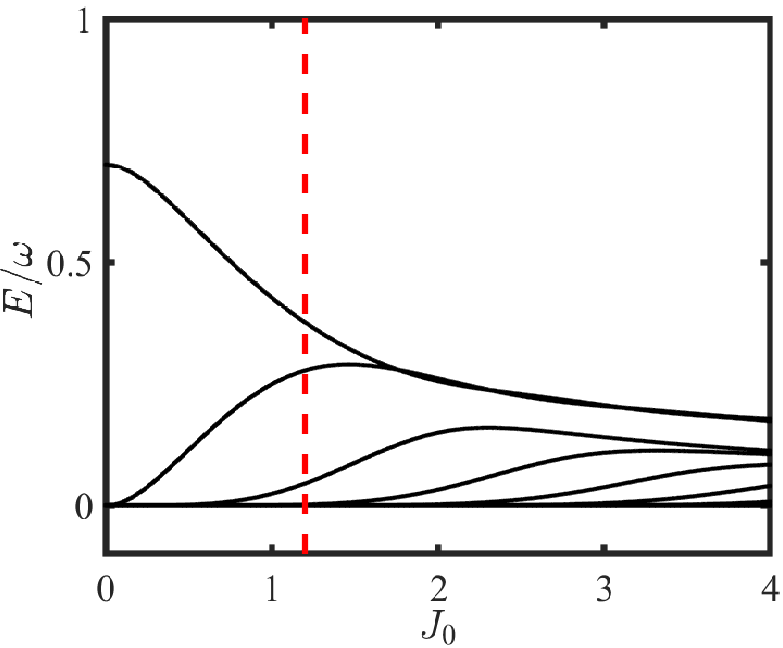}
\caption{(Color online) Quasienergy spectrum $E$ vs. $J_{0}$.
  We diagonalize the Floquet-Bloch lattice model~(\ref{FloquetBlochmodel}) with $\omega=B=1$, $\Delta=0.7$ and $J_1=0.001$.
  The red dashed line corresponds to $J_{0}=1.2$.}
\label{fig:6}
\end{figure}

\begin{figure}[htp]
\center
\includegraphics[width=0.8\textwidth]{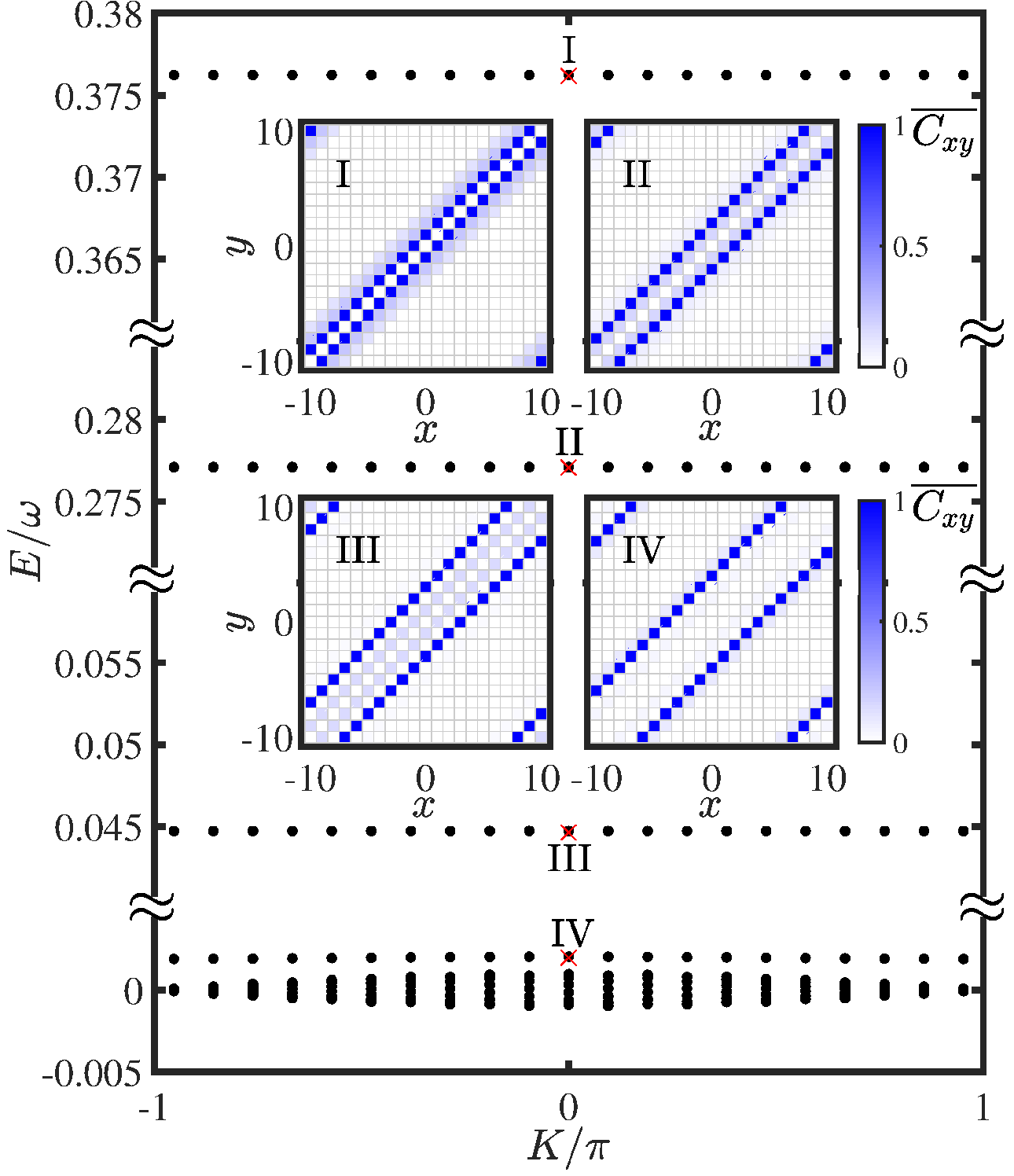}
\caption{(Color online) Quasienergy spectrum $E$ vs. $K$.
  The insets $\rm{\uppercase\expandafter{\romannumeral1}}$, $\rm{\uppercase\expandafter{\romannumeral2}}$, $\rm{\uppercase\expandafter{\romannumeral3}}$ and $\rm{\uppercase\expandafter{\romannumeral4}}$ respectively describe the normalized magnon-magnon correlations $\overline{C_{xy}}=C_{xy}/C^{max}_{xy}$ of Floquet states labeled in the four bound-state band above the continuum band.
  The parameters are chosen as $\omega=B=1$, $\Delta=0.7$, $J_0=1.2$ and $J_1=0.001$.}
\label{fig:7}
\end{figure}

Our system not only is able to induce NNN bound states, but also has potential to engineer longer-range bound states.
In figure~\ref{fig:6}, we show how the quasienergy changes with $J_{0}$.
With the increase of $J_{0}$, more and more modulation-induced bands emerge.
To illustrate how the longer-range bound state is induced, we take $J_{0}=1.2$ marked with red dashed line in figure~\ref{fig:6} as an example.
We obtain the two-body Floquet-Bloch bands versus $K$ by solving the eigenequation~(\ref{eigenequationinKspace}) in figure~\ref{fig:7}.
The parameters are chosen as $\omega=B=1$, $\Delta=0.7$, $J_0=1.2$ and $J_1=0.001$.
Four isolated bands appear above the continuum band in figure~\ref{fig:7}.
We respectively mark the Floquet states from different isolated bands as $\rm{\uppercase\expandafter{\romannumeral1}}$, $\rm{\uppercase\expandafter{\romannumeral2}}$, $\rm{\uppercase\expandafter{\romannumeral3}}$ and $\rm{\uppercase\expandafter{\romannumeral4}}$ in the quasienergy spectrum and analyze their correlated properties in the insets $\rm{\uppercase\expandafter{\romannumeral1}}$, $\rm{\uppercase\expandafter{\romannumeral2}}$, $\rm{\uppercase\expandafter{\romannumeral3}}$ and $\rm{\uppercase\expandafter{\romannumeral4}}$ of figure~\ref{fig:7}.
The magnon-magnon correlations clearly show that, aside from NN and NNN bound states, the bound state with relative distance at three and four sites appear.
It is worth noting that these correlations mostly distributes along their corresponding diagonal lines and partly distribute in other diagonal lines, called as hybridization.
According to our further numerical results, in the appearance of modulation-induced longer-range bound state, hybridization inevitably arises and even more complex case, e.g., the transformation between different types of bound state in the same bound-state band.

~\\
\textbf{{References}}
~\\

\bibliographystyle{apsrev4-1}

\end{document}